\documentclass[prl,aps,twocolumn,showpacs,preprintnumbers,amsmath,amssymb]{revtex4}
\usepackage[dvips]{graphicx}
\usepackage{amsmath}
\usepackage{amssymb}

\usepackage{graphicx}
\usepackage{dcolumn}
\usepackage{bm}
\def\Dsl{\hbox{/\kern-.6700em\it D}} 
\def\dsl{\hbox{/\kern-.5300em$\partial$}}

\def\eqa{\begin{eqnarray}}
\def\eeqa{\end{eqnarray}}
\def\eq{\begin{equation}}
\def\eeq{\end{equation}}
\def\be{\begin{equation}}
\def\ee{\end{equation}}
\def\bea{\begin{eqnarray}}
\def\eea{\end{eqnarray}}

\begin{document}

\preprint{HUTP-05/A0048}

\title{Producing a Scale-Invariant Spectrum of
Perturbations in a Hagedorn Phase of String Cosmology}
\author{Ali Nayeri$^1$}
\email{nayeri@schwinger.harvard.edu}%
\author{Robert H. Brandenberger$^2$}
\email{rhb@karabine.physics.mcgill.ca}%
\author{Cumrun Vafa$^1$}%
\email{vafa@string.harvard.edu}%
\affiliation{\qquad $^1$~Jefferson
Physical Laboratory, Harvard
University,Cambridge, MA 02138, USA \\
$^2$~Department of Physics, McGill University, 3600 University
Street, Montr\'eal QC, H3A 2T8, Canada}

\date{\today}


\begin{abstract}
We study the generation of cosmological perturbations during the
Hagedorn phase of string gas cosmology. Using tools of string
thermodynamics we provide indications that it may be possible
to obtain a nearly scale-invariant spectrum of cosmological
fluctuations on scales which are of cosmological interest today.
In our cosmological scenario, the early Hagedorn phase of string gas
cosmology goes over smoothly into the radiation-dominated phase of
standard cosmology, without having a period of cosmological inflation.
\end{abstract}

\maketitle


{\bf Introduction.} Since superstring theory contains many scalar
fields, it is not unreasonable to study the possibility that a
period of cosmological inflation might naturally arise from string
theory (see e.g. \cite{Linderev,Cliff,JCline} for recent review
articles on this approach). Most of the work on trying to obtain
inflation from string theory, however, is done in the framework of a
low energy effective field theory motivated by string theory, and
does not take into account new symmetries and new degrees of freedom
of string theory which are hard to see at the level of an effective
field theory (see also \cite{cumrun:05} for the restrictions that
string theory can impose on the range of such a scalar field). In
addition, the scalar field-driven inflationary paradigm, although
phenomenologically very successful in terms of predicting an almost
scale-invariant spectrum of adiabatic cosmological perturbations
\cite{pred}, suffers from several conceptual problems (see e.g.
\cite{RHBrev5,RHBrev1} for discussions of these problems). In
particular, there is an initial cosmological singularity. Thus, it
is of great interest to explore the possibility of obtaining a new
paradigm of early universe cosmology which is not based on scalar
field-driven inflation but nevertheless predicts an almost
scale-invariant spectrum of cosmological perturbations.

There is an early approach to string cosmology,
now often called ``string gas cosmology'', which is based specifically
on new symmetries (T-duality) and new degrees of freedom (string
winding modes) of string theory \cite{BV} (see also \cite{Perlt}).
Based on considerations of
string thermodynamics, it was argued that string theory could provide
a nonsingular cosmology. Going backwards in time, the universe
contracts and the temperature grows. However, the temperature will not
exceed the Hagedorn temperature. As radius of space
approaches the self-dual radius (the string length), the pressure of
the string gas will go to zero because of T-duality (the positive
contribution to the pressure from momentum modes will cancel against
the negative pressure from string winding modes). Using the background
equations of motion from dilaton gravity, it follows that the
evolution of the scale factor near the self-dual radius will be
quasi-static \cite{TV}. Once the radius of space decreases below the
self-dual radius, the string gas temperature will decrease,
demonstrating that string gas cosmology will be non-singular.
In addition, a mechanism was proposed \cite{BV} which
might explain why, starting with all spatial dimensions of string
scale, only three spatial dimensions can grow to be macroscopic. There
has recently been quite a lot of work on further developing string gas
cosmology (see e.g. \cite{ABE,Watson,Patil1,Patil2,Edna} and
\cite{RHBrev2,WatBatrev,RHBrev3} for recent reviews and comprehensive lists
of references).

In string gas cosmology, it is assumed that the universe starts in a
Hagedorn phase, a phase in which the universe is quasi-static and in
thermal equilibrium at a temperature close to the Hagedorn temperature
\cite{Hagedorn}, the limiting temperature of perturbative string
theory. As the universe slowly expands, heavy degrees of freedom
gradually fall out of equilibrium. String winding modes keep all but
three spatial dimensions compact \cite{BV} (see, however,
\cite{Columbia,DFM} for a critical view of this aspect of the
scenario).

In this Letter, we will study the generation of cosmological
fluctuations during the early Hagedorn phase of string gas cosmology
using the tools of string statistical mechanics. Since this early
phase is quasi-static, the Hubble radius $H^{-1}(t)$ is very large
(infinite in the limit of the exactly static case). The approximation of
thermodynamic equilibrium is justified on scales smaller than the
Hubble radius. We demonstrate that, in this phase, a gas of closed
strings induces a scale-invariant spectrum of scalar metric
fluctuations on all scales smaller than the Hubble radius. Provided
that the expansion of space is sufficiently slow, these scales will
include all scales which are currently being probed by cosmological
observations
\footnote{We are also assuming by fiat that the three ``large''
  dimensions of space are sufficiently large during the Hagedorn phase
  such that a universe encompassing the presently observed Hubble
  radius can emerge from it during the usual expansion history of
  standard cosmology.}.
Provided that the spectrum in these fluctuations is not distorted at
the time of the transition from the Hagedorn phase to the usual
phase of radiation-domination of standard cosmology (which we argue
is unlikely), it follows that string gas cosmology will lead -
without invoking a period of inflation - to a scale-invariant
spectrum of adiabatic curvature fluctuations.


{\bf Outline of the Analysis.} Figure 1 presents a space-time sketch
of the history of the universe according to string gas cosmology.
The initial phase $t < t_R$ is the Hagedorn phase during which space
is quasi-static, and the temperature is close to the Hagedorn
temperature $T_H$. For times $t
> t_R$, the universe is assumed to be dominated by radiation as in
standard cosmology. Since space-time is quasi-static during the
Hagedorn phase, the Hubble radius is of cosmological scale. During
the transition from the Hagedorn phase to the radiation-dominated
phase, the Hubble radius shrinks dramatically to a microscopic scale
given by
\be H^{-1}(t_R) \, \sim \, \frac{M_{pl}}{T_H^2} \, , \ee
where $M_{pl}$ is the four-dimensional Planck mass. In this paper,
we assume that the radii of the extra spatial dimensions of string
theory have already been stabilized. After $t_R$, the Hubble radius
expands linearly in time \footnote{In order to explain the current
size of the universe, the radius of the three large dimensions must
be at least $10^{23}$ times the string scale at the Hagedorn
temperature - a quantification of the flatness (or size) problem
which our scenario does not resolve.}.

\begin{figure}
\includegraphics[height=8cm]{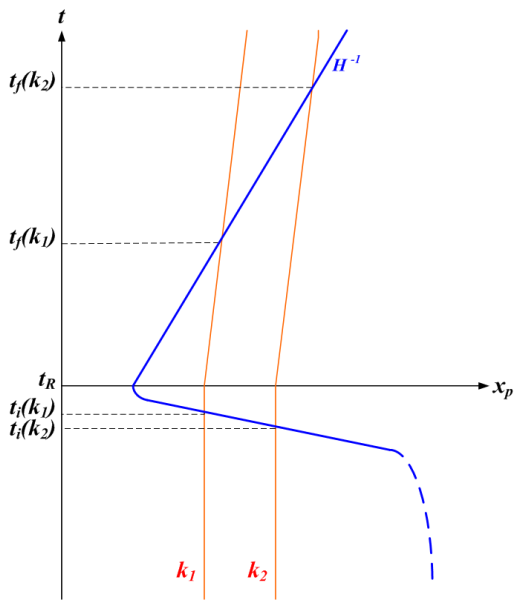}
\caption{Space-time diagram (sketch) showing the evolution of fixed
comoving scales. The vertical axis is time, the horizontal axis is
physical distance. The Hagedorn phase ends at the time $t_R$ and is
followed by the radiation-dominated phase of standard cosmology. The
blue curve represents the Hubble radius $H^{-1}$ which is
cosmological during the quasi-static Hagedorn phase, shrinks
abruptly to a microphysical scale at $t_R$ and then increases
linearly in time for $t > t_R$. Fixed comoving scales (labeled by
$k_1$ and $k_2$) which are currently probed in cosmological
observations have wavelengths which are smaller than the Hubble
radius during the Hagedorn phase. They exit the radius at times
$t_i(k)$ just prior to $t_R$, and propagate with a wavelength larger
than the Hubble radius until they reenter the Hubble radius at times
$t_f(k)$.} \label{fig:1}
\end{figure}

In Figure 1, we also sketch the physical length for various
fluctuation modes $k$ corresponding to fixed comoving wavenumber. The key
feature is that the fluctuations are inside the Hubble radius during
the Hagedorn phase. They exit the Hubble radius at times $t_i(k)$
which are close to the transition time $t_R$. The modes then reenter
the Hubble radius at late times $t_f(k)$.

We want to calculate the amplitude of the metric fluctuations at late
times in the phase of standard cosmology. In order to obtain agreement
with current observations of cosmic microwave (CMB) anisotropies and
large-scale structure, the spectrum of the metric fluctuations
needs to be nearly scale-invariant and nearly adiabatic \cite{WMAP}.
To compute these fluctuations, we will use the theory of linear
cosmological perturbations about a four-dimensional homogeneous and
isotropic cosmology (see \cite{MFB} for a comprehensive review and
\cite{RHBrev4} for a pedagogical introduction).

On scales larger than the Hubble radius, gravity dominates the
dynamics and metric fluctuations play the leading role. We will
calculate here the spectrum of scalar metric fluctuations, fluctuation
modes which couple to the matter sources. In the absence of
anisotropic stress, there is only one physical degree of freedom,
namely the relativistic generalization of the Newtonian gravitational
potential. In longitudinal gauge, the metric then takes the form
\be
ds^2 \, = \, - (1 + 2 \Phi) dt^2 + a(t)^2 (1 - 2 \Phi) d{\bf x}^2 \, ,
\ee
where $t$ is physical time, ${\bf x}$ are the comoving spatial
coordinates of the three large spatial dimensions, $a(t)$ is the
cosmological scale factor and $\Phi({\bf x}, t)$ represents the
fluctuation mode.

On scales smaller than the Hubble radius, the gravitational potential
$\Phi$ is determined by the matter fluctuations via the Einstein
constraint equation (the relativistic generalization of the Poisson
equation of Newtonian gravitational perturbation theory)
\be \label{constr}
\nabla^2 \Phi \, = \, 4 \pi G \delta \rho \, ,
\ee
where $\rho$ is the energy density.

Following the early analyses of the generation and evolution of
cosmological perturbations in inflationary cosmology (see e.g.
\cite{BST,BKP,RHB} for analyses in the spirit of what we do here),
we will in the following calculate the power spectrum of mass
fluctuations on sub-Hubble scales during the Hagedorn phase. We then
use (\ref{constr}) to calculate the magnitude of the metric
fluctuations $\Phi$ at the end of the Hagedorn phase (more
specifically, at the time $t_i(k)$ when the scale labeled by $k$
exits the Hubble radius). As long as the equations of four
space-time dimensional general relativistic cosmological
perturbation theory apply, then $\Phi$ is conserved on super-Hubble
scales as long as the equation of state of the background does not
change significantly \footnote{The use of ordinary general
relativistic cosmological perturbation theory is certainly justified
after the end of the Hagedorn phase, but not necessarily in the time
interval between $t_i(k)$ and $t_R$. We intend to come back to a
careful study of the evolution of fluctuations during this time
interval in a followup paper.}.

Assuming the validity of the arguments of the previous paragraphs,
then the spectral index $n$ of the cosmological perturbations is
determined by
\be
P_{\Phi}(k) \, \equiv \, k^3 |\Phi(k)|^2 \, \sim k^{n - 1} \, .
\ee
In the above, $\Phi(k)$ is the Fourier coefficient of $\Phi$ and $P_X$
denotes the dimensionless
power spectrum of some quantity $X$. The value $n = 1$
corresponds to a scale-invariant spectrum.

Making use of the constraint equation (\ref{constr}), the
dimensionless power spectrum $P_{\Phi}(k)$ of metric fluctuations
can be expressed in terms of the mean square mass perturbations
$\langle(\delta M)^2\rangle$ inside a sphere of radius $R = k^{-1}$:
\be \label{eq5} P_{\Phi} (k) \, = \, 16 \pi^2 G^2 k^{-4}
\langle\bigl( \delta M \bigr)^2\rangle k^6 \, . \ee
where the final factor $k^{-6}$ comes from converting density to
mass. Thus, in order to establish a scale-invariant spectrum, we
need to show that the mean square mass fluctuation $\langle( \delta
M)^2 \rangle$ scales as $k^{-2}$. In the following subsection we
summarize this calculation. For details the reader is referred to a
companion paper \cite{Ali}.


{\bf Computation of the Spectrum.} Now we outline the calculation of
the power spectrum of mass fluctuations. Starting point is the
thermodynamic partition function
\be \label{part}
Z(\beta) \, = \, \sum_i e^{- \beta E_i} \, ,
\ee
where the summation runs over all states, $E_i$ is the energy of the
state, and $\beta$ is the inverse temperature. The assumption here
is the string coupling is sufficiently small, $g_s \ll 1$, and the
local spacetime geometry is close to flat over the length scale of
the finite size box of volume $V = R^d$.  For that we consider a box
of size $H^{-1}$ which consists of $N$ blocks of size $R$.  In each
block, $i$, the universe is homogenous and isotropic and filled with
strings of energy $E_i$. In order to compute the mean square mass
fluctuations in a region of radius $R = k^{-1}$, we apply string
thermodynamics to a volume of that size. From (\ref{part}) we obtain
\be \label{eq7} \langle\bigl( \delta M \bigr)^2\rangle \,  =  \,
\langle E^2\rangle - \langle E \rangle^2 \,  =  \, T^2
\left(\frac{\partial\langle E \rangle}{\partial T}\right)_V \, , \ee
where the angular brackets stand for thermodynamic averaging over
the whole ensemble of the block universes, $\langle E \rangle$ is
the ensemble average energy, and the subscript $V$ indicates that
the partial derivative is taken at constant volume. In terms of the
specific heat $C_V \, \equiv \, \left(\partial \langle E
\rangle/\partial T\right)_V \, , $ the result becomes

\be \langle\bigl( \delta M \bigr)^2\rangle \, = \, T^2 C_V \, . \ee

The specific heat of a gas of closed strings in a background space
given by three large toroidal dimensions and six small compact
dimensions is calculated in a companion paper \cite{Ali}, using the
methods of \cite{Jain}. The result,
evaluated at temperatures close to the Hagedorn temperature, is
\be C_V \, \simeq \, \frac{R^2}{\alpha'^{3/2} T}\frac{1}{1 - T/T_H}
\, , \ee
where $\alpha' = \ell_s^2$, $\ell_s$ being the string length. Note
that it was important to assume that all dimensions are compact in
order to obtain positive specific heat. The mean square mass
fluctuation can now immediately be read off from (\ref{eq7}), and
inserting into (\ref{eq5}) leads to the final result for the
spectrum of metric fluctuations:
\be \label{spectrum} P_{\Phi} \, \sim \, 16 \pi^2 G^2 \alpha'^{-3/2}
\frac{T}{1 - T/T_H} \, \ee
{F}rom our final result (\ref{spectrum}) it follows that the spectrum
of metric fluctuations is approximately scale-invariant, and that
its amplitude is suppressed by the ratio $(\ell_{Pl} / \ell_s)^4$,
where $\ell_{Pl}$ is the four dimensional Planck length. In order to
obtain the observed amplitude of fluctuations \cite{WMAP}, a
hierarchy of lengths of the order of $10^3$ is required. This is
consistent with our initial assumption that the string coupling
constant should be really small since $(\ell_{Pl}/\ell_s) = g_s \sim
10^{-3} \ll 1$. Note that since for fixed value of $k$, the
temperature $T$ which appears in the spectrum is to be evaluated at
the time $t_i(k)$, a slight tilt of the spectrum towards a red
spectrum is induced.


{\bf Discussion and Conclusions.} In this Letter, we have studied
the generation and evolution of cosmological fluctuations in a model
of string gas cosmology in which an early quasi-static Hagedorn
phase is followed by the radiation-dominated phase of standard
cosmology, without an intervening period of inflation. Due to the
fact that the Hubble radius during the Hagedorn phase is
cosmological, it is possible to produce fluctuations using causal
physics. Assuming thermal equilibrium on scales smaller than the
Hubble radius, we have used string thermodynamics to study the
amplitude of density fluctuations during the Hagedorn phase. The
mean square mass fluctuations are determined by the specific heat of
the string gas. To compute the perturbations on a physical length
scale $R$, we apply string thermodynamics to a box of size $R$.
Working under the assumption that all spatial dimensions are compact
(but our three spatial dimensions are sufficiently large), the
specific heat turns out to scale as $R^2$. This is an intrinsically
stringy effect: in the case of point particle thermodynamics, the
specific heat would scale as $R^3$. The $R^2$ scaling of the
specific heat leads to a scale-invariant spectrum of metric
fluctuations.

In order to compute the spectrum of metric fluctuations at late times,
we have applied the usual general relativistic theory of cosmological
perturbations. Whereas this is clearly justified for times $t > t_R$,
its use at earlier times is doubtful. We have used the constraint
equation coming from Einstein gravity to convert the matter
fluctuations into metric perturbations immediately prior to $t_R$,
when scales of cosmological interest today exit the Hubble radius. We
have also assumed that the metric perturbation variable $\Phi$ does
not change on super-Hubble scales during the transition between the
Hagedorn phase and the radiation-dominated phase of standard
cosmology. These assumptions are well justified in the context of the
usual relativistic perturbation theory. However, the fact that the
Hagedorn phase is described by a dilaton gravity background and not by
a purely general relativistic background may lead to some
modifications. However, since it was shown \cite{TV} that the usual
radiation dominated phase with constant dilaton naturally emerges after
the Hagedorn phase, there are good reasons to believe that the changes
to our conclusions will be small.

Although our cosmological scenario provides a new mechanism for
generating a scale-invariant spectrum of cosmological perturbations,
it does not solve all of the problems which inflation solves. In
particular, it does not solve the flatness problem. Without assuming
that the three large spatial dimensions are much larger than the
string scale, we do not obtain a universe which is sufficiently large
today.

Our scenario may well be testable observationally. Taking into account
the fact that the temperature $T$ evaluated at the time $t_i(k)$ when
the scale $k$ exits the Hubble radius depends slightly on $k$, the
formula (\ref{spectrum}) leads to a calculable deviation of the
spectrum from exact scale-invariance. Since $T(t_i(k))$ is decreasing
as $k$ increases, a slightly red spectrum is predicted. Since the
equation of state does not change by orders of magnitude during the
transition between the initial phase and the radiation-dominated phase
as it does in inflationary cosmology, the spectrum of tensor modes is
not expected to be suppressed compared to that of scalar modes. Hence,
a large ratio of tensor to scalar fluctuations might be a specific
prediction of our model. This issue deserves further attention.

\begin{acknowledgments}

{\bf Acknowledgments.} The work of A.N. and C.V. is supported in
part by NSF grant PHY-0244821 and DMS-0244464. R.B. wishes to thank
the Harvard High Energy Theory group for its hospitality during
visits when this work was initiated and completed. The work of R.B.
is supported by funds from McGill University, by an NSERC Discovery
Grant and by the Canada Research Chair program.

\end{acknowledgments}


\begin{thebibliography}{99}

\bibitem{Linderev}
A.~Linde,
  eConf {\bf C040802}, L024 (2004)
  [arXiv:hep-th/0503195].

\bibitem{Cliff}
C.~P.~Burgess,
  Pramana {\bf 63}, 1269 (2004)
  [arXiv:hep-th/0408037].

\bibitem{JCline}
J.~M.~Cline,
  arXiv:hep-th/0501179.

\bibitem{cumrun:05}
C.~Vafa,
  arXiv:hep-th/0509212.

\bibitem{pred}
V.~F.~Mukhanov and G.~V.~Chibisov,
JETP Lett.\  {\bf 33}, 532 (1981)
[Pisma Zh.\ Eksp.\ Teor.\ Fiz.\  {\bf 33}, 549 (1981)];\\
V.~N.~Lukash,
Sov.\ Phys.\ JETP {\bf 52}, 807 (1980) [Zh.\ Eksp.\ Teor.\ Fiz.\
{\bf 79},  (1980)].

\bibitem{RHBrev5}
R.~H.~Brandenberger, [arXiv:hep-ph/9910410].

\bibitem{RHBrev1}
R.~H.~Brandenberger,
  arXiv:hep-th/0509076.

\bibitem{BV}
R.~H.~Brandenberger and C.~Vafa,
  Nucl.\ Phys.\ B {\bf 316}, 391 (1989).

\bibitem{Perlt}
J.~Kripfganz and H.~Perlt,
  Class.\ Quant.\ Grav.\  {\bf 5}, 453 (1988).

\bibitem{TV}
A.~A.~Tseytlin and C.~Vafa,
  Nucl.\ Phys.\ B {\bf 372}, 443 (1992)
  [arXiv:hep-th/9109048].

\bibitem{ABE}
S.~Alexander, R.~H.~Brandenberger and D.~Easson,
  Phys.\ Rev.\ D {\bf 62}, 103509 (2000)
  [arXiv:hep-th/0005212].

\bibitem{Watson}
S.~Watson and R.~Brandenberger,
  JCAP {\bf 0311}, 008 (2003)
  [arXiv:hep-th/0307044].

\bibitem{Patil1}
S.~P.~Patil and R.~Brandenberger,
  Phys.\ Rev.\ D {\bf 71}, 103522 (2005)
  [arXiv:hep-th/0401037].

\bibitem{Patil2}
S.~P.~Patil and R.~H.~Brandenberger,
  arXiv:hep-th/0502069.

\bibitem{Edna}
R.~Brandenberger, Y.~K.~Cheung and S.~Watson,
  arXiv:hep-th/0501032.

\bibitem{RHBrev2}
R. Brandenberger, [arXiv:hep-th/0509099],
  \textit{Challenges for String Gas Cosmology}
  to appear in the proceedings of the 59th Yamada Conference
  ``Inflating Horizon of Particle Astrophysics and Cosmology''
 (University of Tokyo, Tokyo, Japan, June 20 - June 24, 2005).

\bibitem{WatBatrev}
T.~Battefeld and S.~Watson,
  arXiv:hep-th/0510022.

\bibitem{RHBrev3}
R. Brandenberger, \textit{Moduli Stabilization in String Gas
Cosmology}
  to appear in the proceedings of YKIS 2005 (Yukawa Institute for
  Theoretical Physics, Kyoto, Japan, June 27 - July 1, 2005).
  arXiv:hep-th/0509159.

\bibitem{Hagedorn}
R.~Hagedorn,
  Nuovo Cim.\ Suppl.\  {\bf 3}, 147 (1965).

\bibitem{Columbia}
R.~Easther, B.~R.~Greene, M.~G.~Jackson and D.~Kabat,
  JCAP {\bf 0502}, 009 (2005)
  [arXiv:hep-th/0409121].

\bibitem{DFM}
R.~Danos, A.~R.~Frey and A.~Mazumdar,
  Phys.\ Rev.\ D {\bf 70}, 106010 (2004)
  [arXiv:hep-th/0409162].



\bibitem{MFB}
V.~F.~Mukhanov, H.~A.~Feldman and R.~H.~Brandenberger,
  Phys.\ Rept.\  {\bf 215}, 203 (1992).

\bibitem{RHBrev4}
R.~H.~Brandenberger,
  Lect.\ Notes Phys.\  {\bf 646}, 127 (2004)
  [arXiv:hep-th/0306071].

\bibitem{BST}
J.~M.~Bardeen, P.~J.~Steinhardt and M.~S.~Turner,
  Phys.\ Rev.\ D {\bf 28}, 679 (1983).

\bibitem{BKP}
R.~H.~Brandenberger and R.~Kahn,
  Phys.\ Rev.\ D {\bf 29}, 2172 (1984).

\bibitem{RHB}
R.~H.~Brandenberger,
  Nucl.\ Phys.\ B {\bf 245}, 328 (1984).

\bibitem{Ali} A. Nayeri, companion paper.

\bibitem{Jain}
N.~Deo, S.~Jain and C.~I.~Tan,
  Phys.\ Lett.\ B {\bf 220}, 125 (1989);
N.~Deo, S.~Jain and C.~I.~Tan,
  Phys.\ Rev.\ D {\bf 40}, 2626 (1989);
N.~Deo, S.~Jain, O.~Narayan and C.~I.~Tan,
  Phys.\ Rev.\ D {\bf 45}, 3641 (1992).

\bibitem{WMAP}
D.~N.~Spergel {\it et al.}  [WMAP Collaboration],
  Astrophys.\ J.\ Suppl.\  {\bf 148}, 175 (2003)
  [arXiv:astro-ph/0302209].
\end{thebibliography}
\end{document}